\documentclass[twocolumn]{aastex631}
\usepackage{float}
\usepackage{enumitem}

\begin{document}

\title{Searching for Compact Obscured Nuclei in Compton Thick AGN}

\shorttitle{Searching for CONs in CT AGN}
\shortauthors{Johnstone et al.}

\author[0000-0001-7690-3976]{Makoto A. Johnstone}
\affiliation{Department of Astronomy, University of Virginia, 530 McCormick Road, Charlottesville, VA 22903, USA}
\affil{Instituto de Astrof{\'{\i}}sica, Facultad de F{\'{i}}sica, Pontificia Universidad Cat{\'{o}}lica de Chile, Campus San Joaqu{\'{\i}}n, Av. Vicu{\~{n}}a Mackenna 4860, Macul Santiago, Chile, 7820436} 

\author[0000-0003-3474-1125]{George C. Privon}
\affiliation{National Radio Astronomy Observatory, 520 Edgemont Road, Charlottesville, VA 22903, USA}
\affiliation{Department of Astronomy, University of Virginia, 530 McCormick Road, Charlottesville, VA 22903, USA}
\affiliation{Department of Astronomy, University of Florida, P.O. Box 112055, Gainesville, FL 32611, USA}

\author[0000-0003-0057-8892]{Loreto Barcos-Munoz}
\affiliation{National Radio Astronomy Observatory, 520 Edgemont Road, Charlottesville, VA 22903, USA}
\affiliation{Department of Astronomy, University of Virginia, 530 McCormick Road, Charlottesville, VA 22903, USA}

\author[0000-0003-2638-1334]{A.S. Evans}
\affiliation{Department of Astronomy, University of Virginia, 530 McCormick Road, Charlottesville, VA 22903, USA}
\affiliation{National Radio Astronomy Observatory, 520 Edgemont Road, Charlottesville, VA 22903, USA}

\author[0000-0002-5828-7660]{S. Aalto}
\affil{Department of Space, Earth and Environment, Onsala Space Observatory, Chalmers University of Technology, 43992 Onsala, Sweden}

\author[0000-0003-3498-2973]{Lee Armus}
\affil{IPAC, California Institute of Technology, 1200 East California Boulevard, Pasadena, CA 91125, USA}

\author[0000-0002-8686-8737]{Franz E. Bauer}
\affil{Instituto de Astrof{\'{\i}}sica, Facultad de F{\'{i}}sica, Pontificia Universidad Cat{\'{o}}lica de Chile, Campus San Joaqu{\'{\i}}n, Av. Vicu{\~{n}}a Mackenna 4860, Macul Santiago, Chile, 7820436} 
\affiliation{Millennium Institute of Astrophysics, Nuncio Monse{\~{n}}or S{\'{o}}tero Sanz 100, Of 104, Providencia, Santiago, Chile}

\author[0000-0002-2183-1087]{L. Blecha}
\affil{Department of Physics, University of Florida, 2001 Museum Rd., Gainesville, FL 32611, USA}

\author[0000-0001-8608-0408]{J. S. Gallagher}
\affil{Department of Astronomy, University of Wisconsin-Madison, 475 N Charter Street, Madison, WI 53706, USA}
\affil{Department of Physics and Astronomy, Macalester College, 1600 Grand Abe, St. Paul, MN 55105, USA}

\author[0000-0001-6174-8467]{S. K\"onig}
\affil{Department of Space, Earth and Environment, Onsala Space Observatory, Chalmers University of Technology, 43992 Onsala, Sweden}

\author[0000-0001-5231-2645]{Claudio Ricci}
\affil{Instituto de Estudios Astrofísicos, Facultad de Ingeniería y Ciencias, Universidad Diego Portales, Avenida Ejercito Libertador 441, Santiago, Chile}
\affil{Kavli Institute for Astronomy and Astrophysics, Peking University, Beijing 100871, People’s Republic of China}

\author[0000-0001-7568-6412]{Ezequiel Treister}
\affil{Instituto de Astrof{\'{\i}}sica, Facultad de F{\'{i}}sica, Pontificia Universidad Cat{\'{o}}lica de Chile, Campus San Joaqu{\'{\i}}n, Av. Vicu{\~{n}}a Mackenna 4860, Macul Santiago, Chile, 7820436} 

\newcommand{\NRAO}{\affiliation{National Radio Astronomy Observatory, 520 Edgemont Road, Charlottesville, VA 22903, USA}}
\author[0000-0002-1185-2810]{Cosima Eibensteiner}
\altaffiliation{Jansky Fellow of the National Radio Astronomy Observatory}
\NRAO

\author[0000-0001-6527-6954]{Kimberly L. Emig}
\affil{National Radio Astronomy Observatory, 520 Edgemont Road, Charlottesville, VA 22903, USA}

\author[0009-0002-2049-9470]{Kara N. Green}
\affiliation{Department of Astronomy, University of Virginia, 530 McCormick Road, Charlottesville, VA 22903, USA}

\author[0000-0002-1568-579X]{Devaky Kunneriath}
\affil{National Radio Astronomy Observatory, 520 Edgemont Road, Charlottesville, VA 22903, USA}

\author[0009-0002-6248-3688]{Jaya Nagarajan-Swenson}
\affiliation{Department of Astronomy, University of Virginia, 530 McCormick Road, Charlottesville, VA 22903, USA}

\author[0000-0003-4546-3810]{Alejandro Saravia}
\affiliation{Department of Astronomy, University of Virginia, 530 McCormick Road, Charlottesville, VA 22903, USA}

\author[0000-0001-9163-0064]{Ilsang Yoon}
\affil{National Radio Astronomy Observatory, 520 Edgemont Road, Charlottesville, VA 22903, USA}

\begin{abstract}

Compact Obscured Nuclei (CONs) are
heavily obscured infrared cores that have been found in local (ultra)luminous infrared galaxies (U/LIRGs). They show bright emission from vibrationally excited rotational transitions of HCN, known as HCN-vib, and are thought to harbor Compton Thick (CT, $N_{\text{H}} \geq 10^{24}$ cm$^{-2}$) active galactic nuclei (AGN) or extreme compact starbursts. We explore the potential evolutionary link between CONs and CT AGN by searching for CONs in hard X-ray-confirmed CT AGN from the Great Observatories All-sky LIRG Survey (GOALS). Here, we present new Atacama Large Millimeter/submillimeter Array Band 6 observations that targeted HCN-vib emission in four hard X-ray-confirmed CT AGN. We analyze these objects together with literature HCN-vib measurements of five additional hard X-ray-confirmed CT AGN from the GOALS sample. We do not detect any CONs in this combined sample of nine CT AGN. We then explore a proposed evolutionary sequence in which CONs evolve into X-ray-detectable CT AGN once outflows and feedback reduce the column densities of the enshrouding gas. We find, however, no evidence of well-developed dense molecular outflows in the observed CT AGN. While this could suggest that CT AGN are not universally linked to CONs, it could also be explained by a short duty cycle for molecular outflows.
\end{abstract}

\section{Introduction} \label{sec:intro}

Recent Atacama Large Millimeter/submillimeter Array (ALMA) observations have revealed deeply embedded hot (T$_{\text{dust}} >$ 100 K) infrared cores ($r<100$ pc) in some local (ultra) luminous infrared galaxies (U/LIRGs) \citep{Sakamoto2010,Imanishi2013,Aalto2015, Martin2016, Falstad2021}. These compact obscured nuclei (CONs) are characterized by a strong mid-infrared radiation field that is formed via the radiative trapping of 14$\mu$m continuum photons by large molecular gas columns ($N_{H} > 10^{24}$ cm$^{-2}$; \citealt{GA2019, Falstad2021}). Such radiation fields could be powered by either nuclear starbursts or active galactic nuclei (AGN) \citep{Andrews2011,Aalto2015,GA2019}, but the exact power source is unknown. Unveiling the central engine of CONs thus remains a fundamental question in understanding the nature of these infrared cores. A starburst-scenario would require an extreme top-heavy initial mass function (IMF), consisting of mostly massive O-stars \citep{Aalto2019}. An AGN-scenario, on the other hand, implies the presence of a Compton Thick (CT, $N_{\text{H}} \geq 10^{24} \text{ cm}^{-2}$) AGN due to the obscured nature of these systems \citep{Aalto2015,Aalto2019}. 

The mechanisms that drive the build  up of this nuclear obscuration are still unclear. CONs have been identified in major mergers, minor mergers, and isolated galaxies \citep{Falstad2021}, suggesting that both internal and interaction-driven processes can be responsible for their extreme column densities \citep{Falstad2021}. There are, however, striking differences in the rate of occurrence of CONs based on the infrared luminosity of the host galaxy \citep{Falstad2021}. In fact, \cite{Falstad2021} found that $\sim$40\% of local ULIRGs ($L_{\text{IR}} \geq 10^{12}$ $L_{\odot}$) and $\sim$20\% of local LIRGs ($10^{11}$ $_L{\odot} \leq$ $L_{\text{IR}} < 10^{12}$ $L_{\odot}$) host CONs, whereas no CONs have been identified in local lower luminosity galaxies ($L_{\text{IR}} < 10^{11}$ $L_{\odot}$). This luminosity-dependence makes U/LIRGs the only known hosts of the CON phenomenon.

Optically thick layers of dust and gas pose an observational challenge when studying CONs. Interpreting spectral line emission often proves to be difficult due to absorption from foreground layers of obscuring gas (e.g. \citealt{Aalto2015, Aalto2019}). The greenhouse effects of the H$_2$ columns, however, create a high column density ($N_{\text{H}} > 2 \times 10^{23}$ cm$^{-2}$) and high brightness temperature ($T_b \sim 100$ K) environment that can vibrationally excite ($v_2=1f$) rotational transitions of HCN (3-2) (hereafter, HCN-vib) via radiative pumping \citep{Ziurys1986, GA2019}. Bright HCN-vib emission, therefore, indicates the presence of a CON-like infrared core, making it a crucial tool in identifying and characterizing CONs.

Though it is unclear whether CONs harbor CT AGN, there are uncanny similarities in their properties. In addition to their similarly high column densities ($N_{\text{H}} > 10^{24}$ cm$^{-2}$; \citealt{GA2019}) and high bolometric luminosities \citep{GA2019, Ricci2021}, both CONs and CT AGN have high rates of occurrence in local U/LIRGs \citep{Ricci2017b, Ricci2021, Falstad2021}. It is therefore possible that CONs and CT AGN are manifestations of the same phenomenon (that is, an obscured actively-growing SMBH), but their relationship is yet to be systematically studied. 

Mechanical feedback-based theories suggest that the outflow properties of CONs may give insight into their connection to CT AGN. For example, though many U/LIRGs have fast far-infrared OH outflows \citep{Veilleux2013}, CONs are notably missing them \citep{Falstad2019}.  Instead, they host compact molecular outflows that have been detected at sub-millimeter wavelengths \citep{GB2015, PS2016, PS2018, BM2018, Falstad2018, Fluetsch2019, Lutz2020, Falstad2021}. At least three of these molecular outflows have confirmed collimated morphologies (Arp 220W \citep{BM2018}, ESO 320-G030 \citep{PS2016, Gorski2024}, and Zw 049.057 \citep{Falstad2018}) that are reminiscent of the molecular jets seen in young stellar objects (e.g, \citealt{Plunkett2015}). 

These unique properties motivate the theory that CONs harbor hidden CT AGN that evolve into X-ray detectable CT AGN as their active feedback mechanisms develop. The high column densities of CONs generally preclude the X-ray detection of any AGN that might be present. As their molecular outflows evolve and widen, however, they would sweep or erode away a fraction of the obscuring gas, creating lower column density sight-lines through which the CON-like mid-infrared radiation field would leak out \citep{GA2017b, Falstad2019, Falstad2021}. Once the line of sight column density has been reduced to orders of $N_{\text{H}} = 10^{24} - 10^{25}$ cm$^{-2}$, the CT AGN would be X-ray detectable, but the greenhouse-like conditions necessary to excite HCN-vib molecules may no longer be present \citep{GA2019}. If the AGN bolometric luminosity remains high, the molecular outflow is expected to continue its expansion or even increase in velocity due to the loss of mass in the nuclear interstellar medium which was previously slowing the outflow's expansion. CT AGN following this proposed sequence may be missing CON-like HCN-vib emission, but they would have well-developed molecular outflows. In this work, we investigate this evolutionary scenario by searching for evolved molecular outflows in CT AGN. 

\begin{deluxetable*}{ccccccc}[!ht]
\tablecaption{Compton Thick AGN Studied in This Work \label{tab:sample}}
\tablehead{ 
\colhead{$IRAS$ Name} & \colhead{Source} & \colhead{$z$}& \colhead{$D_L$}  & \colhead{log($L_{\text{IR}}/$$L_\odot$)} & \colhead{log$N_{\text{H}}$} & \colhead{$f_{25}/f_{60}$}  \\[-0.2cm] 
\colhead{} & \colhead{} & \colhead{} & \colhead{(Mpc)}  & \colhead{} & \colhead{(cm$^{-2}$)} & \colhead{} } 

\startdata
F02401--0013 & NGC 1068 & 0.0038 & 15.9 & 11.40 & $\geq$ 24.99 & 0.448 \\
F12590+2934 & NGC 4922N$^*$ & 0.0232 & 111.0 & 11.38 & 25.10 [24.63–NC] & 0.195  \\
13120--5453 & -- &  0.0308 & 144.0 & 12.32 & 24.50 [24.27-24.74] & 0.072 \\
F13229--2934 & NGC 5135 & 0.0137 & 60.9 &  11.30 & 24.80 [24.51-25.00] & 0.141 \\
F16504+0228  & NGC 6240N & 0.0245 &  116.0 & 11.93 & 24.19 [24.09-24.36] & 0.155 \\
F16504+0228 & NGC 6240S & 0.0245 & 116.0 & 11.93 & 24.17 [24.11-24.23] & 0.155 \\
20264+2533 & NGC 6921$^*$ & 0.0139 & 64.2 & 11.11 & 24.15 [23.83-24.40] & 0.358 \\
F21453--3511 & NGC 7130$^*$ & 0.0162 &  72.7 & 11.42 & 24.61 [24.50-24.66] & 0.055 \\
F23254+0830 & NGC 7674$^*$ & 0.0289 & 125.0 & 11.56 & $\geq$ 24.48 & 0.129 
\enddata
\tablenotetext{}{NC: value not constrained.} 
\tablenotetext{*}{New observations are presented in this work.}
\tablecomments{Column (1): Source name. Column (2): Redshifts from \cite{Armus2009}. Column (3): Luminosity distances from \cite{Armus2009}. Column (4): Infrared luminosities from \cite{Armus2009}. Column (5): Line of sight column densities from \cite{Ricci2021}. Column (6): Ratio of IRAS fluxes at 25$\mu$m and 60$\mu$m from \cite{Sanders2003}.}
\end{deluxetable*}
\vspace{-0.8cm}

In this paper, we analyze a sample of nine hard X-ray-confirmed CT AGN; we present four new Band 6 ALMA HCN-vib observations and use archival measurements for five additional sources. We study the relationship between CT AGN and CONs, their distinctive properties, and their potential evolutionary connections. In Section \ref{sec:obs}, we report the sample selection process, the observational set-up, and the data reduction procedure. In Section \ref{sec:results}, we examine the  HCN-vib line intensity and the kinematics of the dense gas tracers, HCN (3-2) and HCO$^+$ (3-2). In Section \ref{sec:CT_CON}, we discuss the potential connection between CT AGN and CONs. We conclude our findings in Section \ref{sec:concl}. 

\section{Sample and Observations} \label{sec:obs} 

\subsection{Sample}

We study the prevalence of CONs in CT AGN ($N_{\text{H}} \geq 10^{24} \text{ cm}^{-2}$) by analyzing observations of HCN-vib (rest frequency = 267.199 GHz for $J= 3-2$ $v_2=1f$). Since the CON phenomenon has only been identified in U/LIRGs and appears to have a strong luminosity-dependence, we specifically target CT AGN in U/LIRGs. In the Great Observatories All-sky LIRG Survey (GOALS) sample \citep{Armus2009}, 14 CT AGN have been identified via \textit{NuSTAR} hard X-ray observations \citep{Bauer2015,Teng2015,Ricci2016,Oda2017,Gandhi2017,Ricci2017a,Iwasawa2020,Yamada2020, Ricci2021}. Of these, five had pre-existing millimeter-wave ALMA observations of HCN-vib (NGC 1068, IRAS 13120-5453, NGC 5135, NGC 6240N, NGC 6240S; \citealt{Privon2017, Imanishi2016, Imanishi2020, Falstad2021, Nishimura2024}). We also conducted new ALMA observations of four additional CT AGN from the GOALS sample (NGC 4922N, NGC 6921, NGC 7130, NGC 7674). These objects were selected due to their visibility to ALMA (dec. $<$ 40 deg) and lack of prior HCN-vib observations. Combined with the pre-existing observations, this results in a total sample of nine CT AGN. The properties of these nine sources are reported in Table \ref{tab:sample}. 

Previous studies have used the ratio of the HCN-vib luminosity to the total infrared luminosity of the host galaxy to identify CONs with a fiducial value of $L_{\text{HCN-vib}}$/$L_{\text{IR}} >$ 10$^{-8}$ \citep{Aalto2015,Falstad2019}. \cite{Falstad2021}, however, suggested that this criterion may be biased against galaxies with a spatially-extended infrared emission region or multiple nuclei. In accordance with \cite{Falstad2021}, we adopt an alternative fiducial value of an HCN-vib surface brightness $\Sigma_{\text{HCN-vib}} \geq$ 1 $L_\odot$ pc$^{-2}$ over a spatial region of $r > 5$ pc as the detection criterion for CONs. This fiducial value is purely empirical and was selected for observations with angular resolutions of $0.2"-0.8"$ ($\sim5-100$ pc; \citealt{Falstad2021}). Note that while our analysis emphasizes this $\Sigma_{\text{HCN-vib}}$ definition for CONs, we return to consider the $L_{\text{HCN-vib}}$/$L_{\text{IR}}$ for completeness in Section \ref{subsubsec:res}.

\begin{figure*}[!ht]
    \centering
    \includegraphics[width=0.85\textwidth]{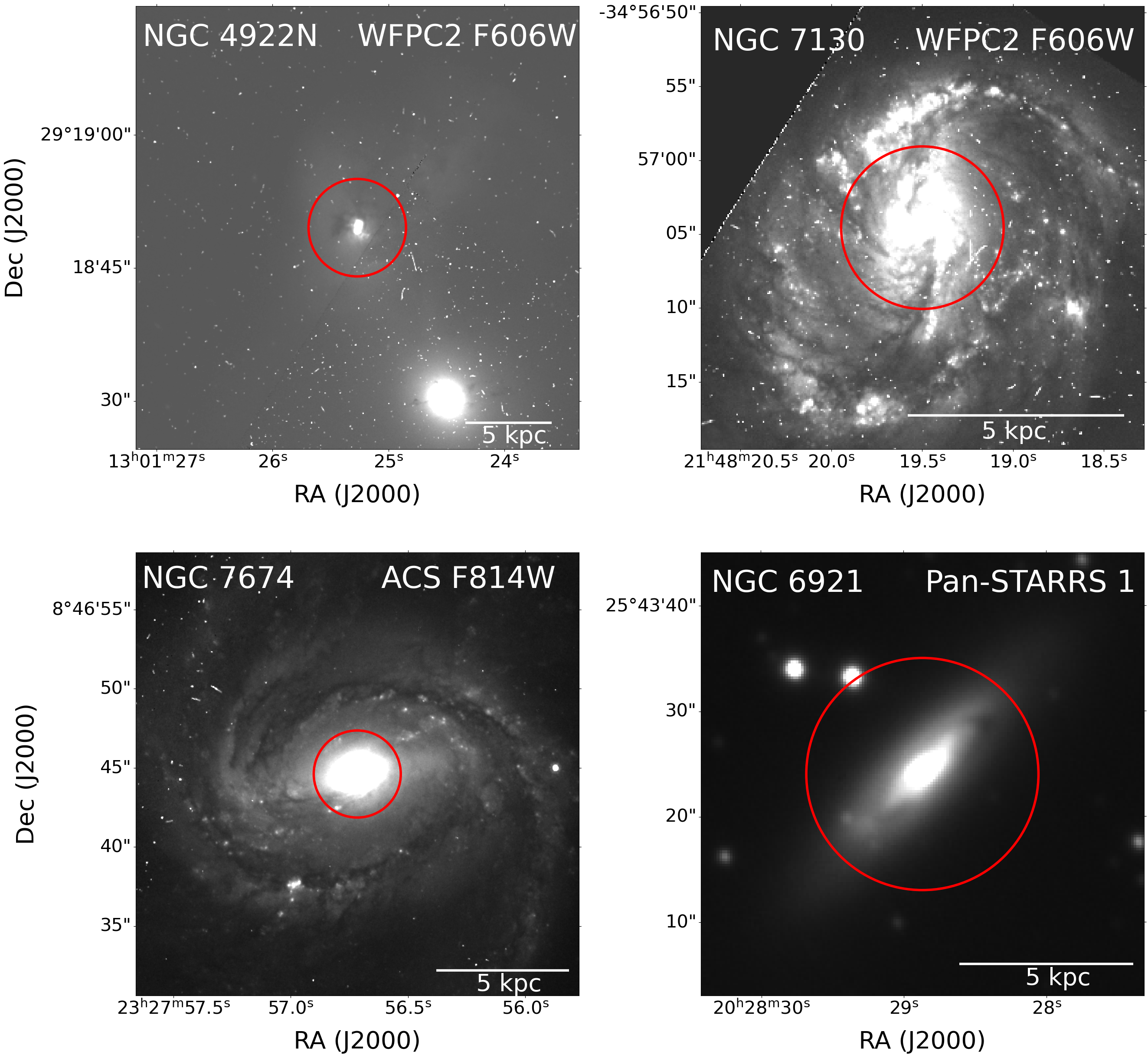}
    \caption{Optical \textit{Hubble Space Telescope} (HST) or PanSTARRS-1 images of the four CT AGN in local U/LIRGs with new observations in our sample. The approximately Band 6 ALMA field of view ($\sim$22$"$) is overlaid in red. HST images are from the following proposal IDs: 5479 (NGC 4922N, NGC 7130) and 10592 (NGC 7674). The NGC 6921 Pan-STARRS 1 image is from project ID 1918. }
    \label{fig:optical}
\end{figure*}

\subsection{New ALMA observations}

New ALMA 12-m Array Band 6 observations of four CT AGN (NGC 4922N, NGC 6921, NGC 7130, NGC 7674) were conducted as a part of Project \#2017.1.00598.S (PI: Privon, G.C.). The approximate ALMA Band 6 field of view ($\sim$ 22$"$) of these observations are shown as a red circle in Figure\,\ref{fig:optical}, overlaid on optical images of the observed galaxies. These targets were observed in the C43-4 and C43-5 ALMA configurations (maximum resolvable scales of 4.3$"$ and 2.4$"$, respectively). Table \ref{tab:obs} summarizes the observational set-up. Atmospheric conditions were reasonable for all observations (T$_{\text{sys}}$ $\leq$ 100 K). Two spectral windows were observed for each source, each centered at the redshifted frequencies of HCN (3-2) and HCO$^+$ (3-2). We adopt the rest frequencies 265.886 GHz and 267.558 GHz for HCN (3-2) and HCO$^+$ (3-2) respectively. These rest frequencies were obtained from the astronomical spectroscopy database, Splatalogue.\footnote{\url{https://splatalogue.online/sp_basic.html}} Spectral windows were set with 1.875 GHz bandwidth and 7.8 MHz spectral resolution. 

\begin{deluxetable*}{ccccccccc}[!ht]
\tablecaption{ALMA observational set-up and image properties\label{tab:obs}}
\tablehead{ \colhead{Source} & \colhead{Date of}  & \colhead{$t_{\text{on}}$} & \colhead{Configuration} & \colhead{$B_{\text{min}}$/$B_{\text{max}}$}  & \colhead{Beam Size} & \colhead{Beam Size} & \colhead{Cube RMS} & \colhead{Continuum RMS} \\[-0.2cm] 
\colhead{} & \colhead{observations} & \colhead{(min)} & \colhead{} & \colhead{(m/km)} & \colhead{($" \times ", ^{\circ}$)} & \colhead{(pc $\times$ pc)} 
 & \colhead{(mJy beam$^{-1}$)} & \colhead{(mJy beam$^{-1}$)}} 

\startdata
NGC 4922N & 2018 Sep 18 & 38.32 & C43-5 & 15/1.397 & 0.38$\times$0.24, 16$^{\circ}$ & 188$\times$119 & 0.25 & 0.06\\
\hline
NGC 7130 & 2018 Mar 25  & 5.13 & C43-4 & 14/0.783 & 0.32$\times$0.29, 71$^{\circ}$ & 109$\times$99 & 0.50 & 0.17 \\
& 2018 Sep 9  & 5.13 & C43-4 & 14/0.783 & -- & -- & -- & -- \\
& 2018 Sep 24  & 5.12 & C43-5 & 14/1.397 & -- & -- & -- & -- \\
\hline 
NGC 7674 & 2018 Sep 19  & 44.97 & C43-5 & 15/1.397 & 0.28$\times$0.24, 54$^{\circ}$ & 167$\times$145 & 0.19 & 0.07 \\
\hline
NGC 6921 & 2018 Sep 14  & 32.18 & C43-5 & 14/1.231 & 0.39$\times$0.27, 12$^{\circ}$ & 115$\times$80 & 0.20 & 0.09 \\
& 2018 Sep 15 & 32.17 & C43-5 & 14/1.261  & -- & -- & -- & -- \\
\enddata

\tablecomments{Column (1): Source name. Column (2): Date of observations. Column (3): On-source observation time. Column (4): Approximate observing configuration. Column (5): Minimum and maximum baselines. Column (6): Synthesized beam size in arcseconds with position angle in degrees. Column
(7): Projected synthesized beam size in parsecs. Column (8): RMS sensitivity of data cubes, defined as mean 1$\sigma$ noise level per 33 km s$^{-1}$ channel. Column (9): RMS sensitivity of continuum images, defined as mean 1$\sigma$ noise level. Properties in Columns 7 through 9 are based on combined imaging for all execution blocks for each source. }
\end{deluxetable*}
\vspace{-0.8cm}

Data reduction, imaging, and analysis were carried out using the Common Astronomy Software Application (CASA, \citealt{CASA2022}) package version 5.1.1-5. Continuum-subtracted visibilities were created using the automated ALMA imaging pipeline \citep{Hunter2023}. The NGC 6921 data set, however, was reprocessed with a manually specified frequency range for the continuum fitting and subtraction because the pipeline inadvertently included some line emission in the baseline fit. To improve the signal-to-noise ratio, the data cubes were smoothed to $\sim$ 33 km s$^{-1}$ channel widths-- approximately four times that of the observatory-provided pipeline products. Data were imaged with Briggs weighting (robust = 0.5;  \citealt{Briggs1995}), and resolutions of $\sim$ 0.3" were achieved. 1-millimeter continuum images were also created from the spectral line-free channels.  Table \ref{tab:obs} includes a summary of the beam sizes and the root-mean-square (RMS) sensitivities of these new ALMA observations. 

\subsection{Literature HCN-vib Measurements}

From the literature, we obtain HCN-vib surface brightness measurements and upper limits of five CT AGN in U/LIRGs (NGC 1068, IRAS 13120-5453, NGC 5135, NGC 6240N, NGC 6240S). For NGC 1068, a $\sim5\sigma$ HCN-vib detection is reported at its western peak at $\sim2$ pc-scales \citep{Imanishi2020}, but HCN-vib non-detections and upper-limits are reported at $10-15$ pc-scales \citep{Imanishi2016,Falstad2021}. In this work, we use the HCN-vib upper limit reported in \citet{Falstad2021} at $\sim15$ pc-scales, as our fiducial value requires measurements over a region with a radius of at least 5 pc. This also ensures that the HCN-vib measurements were taken over similar angular scales across all sources. For NGC 5135 and IRAS 13120-5453, HCN-vib upper limits are presented in \cite{Falstad2021}. HCN-vib non-detections are also reported for NGC 6240N and NGC 6240S in \cite{Nishimura2024}. For the two nuclei, we calculate $3\sigma$ surface brightness upper-limits using the reported HCN (3-2) spectral line properties and data cube RMS, assuming a boxcar profile and a line width equal to the full-width at half-maximum (FWHM) of the detected HCN (3-2) line. These HCN-vib measurements and their implications are presented and discussed in Section \ref{sec:CT_CON}.

\section{Results}  \label{sec:results}

\subsection{1-millimeter continuum }

\begin{figure*}[!ht]
\centering
\includegraphics[width=\textwidth]{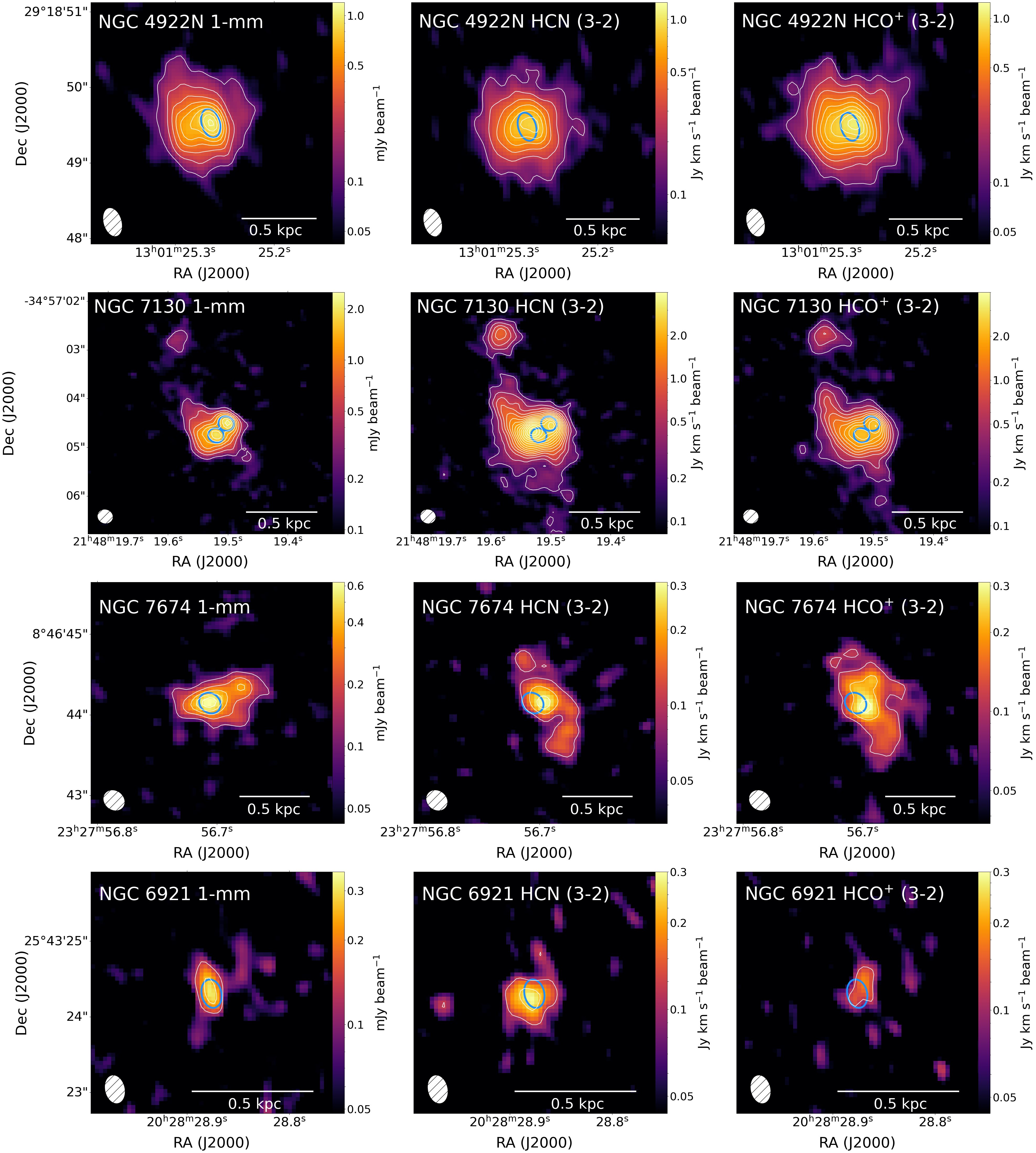}
\caption{1-mm continuum and total intensity (moment 0) maps of HCN (3-2) and HCO$^{+}$ (3-2). Each row corresponds to a source (from top to bottom: NGC 4922N, NGC 7130, NGC 7674, and NGC 6921). For each source, the 1-mm continuum image (left) and moment 0 maps for HCN (3-2) (middle) and HCO$^{+}$ (3-2) (right) are presented. White contours correspond to 4$\sigma$ levels (4$\sigma$, 8$\sigma$, 12$\sigma$...). ALMA synthesized beam sizes are shown in the bottom left of each map. Blue ellipses  correspond to spectral extraction regions for constraining HCN-vib. Note that the images are shown on a logarithmic color-scale with the lower-limit set to $1\sigma$. \label{fig:mom0}}
\end{figure*}

Continuum emission at 1 millimeter (1-mm) was detected and spatially-resolved in all four newly observed CT AGN. Their morphology is shown in Figure \ref{fig:mom0}. We assume the peak of the 1-mm continuum to be the AGN position where HCN-vib emission is most likely to be detected. For this reason, we also assign the 1-mm continuum peak to be the position at which we extract the spectral line profiles (indicated as blue ellipses in Figure \ref{fig:mom0}). 

We note that the NGC 7130 continuum is double-peaked with a projected separation of $d_{\text{sep}} \approx 250$ pc. \cite{Zhao2016} proposed that the AGN position likely corresponds with the eastern peak due to its alignment with the Very Large Array 8.4 GHz continuum detection. Interestingly, we find that the western peak is brighter in the 1-mm continuum (Figure \ref{fig:mom0}). Though AGN-dominated emission is expected to have a flat mm-wave slope ($\alpha_{\text{mm}} \approx 0$, \citealt{Kawamuro2022} where $S \propto \nu^\alpha$), the low sensitivity and the limited bandwidth of the observed spectral windows results in spectral index uncertainties as high as $\sim\pm$ 20, making it difficult to constrain a precise in-band spectral index. Since further observations are necessary to constrain the AGN position, we present results for both continuum peaks in this paper. 

NGC 7674 also has a secondary peak to the northwest ($d_{\text{sep}} \approx$ 105 pc). This secondary peak is spatially coincident with the cm-wave steep spectral-index ($\alpha_{\text{cm}} \approx -1$) synchrotron-dominated radio jet identified in \cite{Song2022}. The brighter primary peak, in contrast, is spatially coincident with AGN-dominated flat-spectrum ($\alpha_{\text{cm}} \approx 0$) radio emission \citep{Song2022}. We therefore assume that the AGN position corresponds to the brighter primary peak for the remainder of this paper. 

We measure continuum properties via a two-dimensional Gaussian fit to the continuum map (using CASA task \texttt{IMFIT}) for consistency with \cite{Falstad2021}. The fit was performed over a region that contained all continuum emission $>3\sigma$. For NGC 7130, two Gaussian components were fitted to account for significant flux contribution from the secondary peak. The peak positions of these two components were fixed to the observed continuum peaks to ensure they were tracing a physical phenomenon. Setting them as free parameters, in contrast, resulted in residual maps with excess flux at both peaks, indicating a poorer fit. Uncertainty analysis of two-dimensional Gaussian-fitting method can be found in \cite{Condon1997}. The measured continuum properties are reported in Table \ref{tab:cont}.

\subsection{Dense gas tracers}

The high dipole moments and high critical densities of HCN (3-2) and HCO$^+$ (3-2) ($n_{\text{crit}} = 10^7$ cm$^{-3}$ and $3 \times 10^6$ cm$^{-3}$, respectively) make them excellent tracers of high-density gas in galaxies. Here, we study the morphology of the dense gas emission. 

\begin{deluxetable*}{cccccc}[!ht]
\footnotesize
\tablecaption{Gaussian-fitted continuum properties \label{tab:cont}}
\tablehead{ \colhead{Source} & \colhead{R.A. of Peak} & \colhead{Dec. of Peak} & \colhead{Flux Density}  & \colhead{Continuum Size} &   \colhead{$S_{\text{1-mm}}$/$\Omega$}\\[-0.2cm] 
\colhead{} &  \colhead{(J2000)} & \colhead{(J2000)} &  \colhead{(mJy)} &  \colhead{(mas$\times$mas)}  & \colhead{(mJy arcsec$^{-2}$)}} 
\startdata
NGC 4922N & 13h 01m 25.265s & +29d 18m 49.55s & 4.50 $\pm$ 0.23 & 738 $\pm$ 33 $\times$ 673 $\pm$ 30 & 8.00 $\pm$ 0.65 \\
NGC 7130 & 21h 48m 19.501s$^*$  & $-$34d 57m 04.50s$^*$ & 2.18 $\pm$ 0.35 & 376 $\pm$ 41 $\times$ 317 $\pm$ 30 & 26.34 $\pm$ 3.23 \\
& 21h 48m 19.520s$^*$ & $-$34d 57m 04.80s$^*$ & 8.89 $\pm$ 0.92 & 686 $\pm$ 79 $\times$ 589 $\pm$ 70 & 17.62 $\pm$ 3.33 \\
NGC 7674 & 23h 27m 56.704s & +08d 46m 44.18s & 2.32 $\pm$ 0.17 & 762 $\pm$ 52 $\times$ 399 $\pm$ 22 & 6.73 $\pm$ 0.77 \\
NGC 6921 & 20h 28m 28.876s & +25d 43m 24.33s & 0.48 $\pm$ 0.09 & 550 $\pm$ 81 $\times$ 287 $\pm$ 25 & 2.68 $\pm$ 0.68
\enddata
\tablenotetext{*}{Fixed parameter}
\tablecomments{Column (1): Source name. Column (2): J2000 Right Ascension of the observed 1-mm continuum peak position(s). Column (3): J2000 Declination of the observed 1-mm continuum peak position(s). Column (4): Flux density in mJy. Column (5): Convolved major and minor axes of the 1-mm continuum in milliarcseconds. Column (6): 1-mm continuum surface brightness. }
\vspace{-0.5cm}
\end{deluxetable*}
\vspace{-0.8cm} 

We detect HCN (3-2) and  HCO$^{+}$ (3-2) emission in all of the four observed CT AGN. We generate total intensity (moment 0) maps of the two spectral lines by integrating over all channels with $>3\sigma$ line emission detections using the Python package \texttt{Spectral-Cube}. These moment 0 maps are shown in Figure \ref{fig:mom0}. Notably, the morphology of the dense gas emission in NGC 7130 and NGC 7674 appears consistent with the optical spiral arms seen in Figure \ref{fig:optical}. NGC 4922N and NGC 6921, however, have more centralized emission that are also consistent with the optical images (Figure \ref{fig:optical}).  Furthermore, the continuum and dense gas emission appear to have broadly similar morphologies in all sources, with the exception of NGC 7674. 

\begin{figure*}[!hp]
\includegraphics[width=\textwidth]{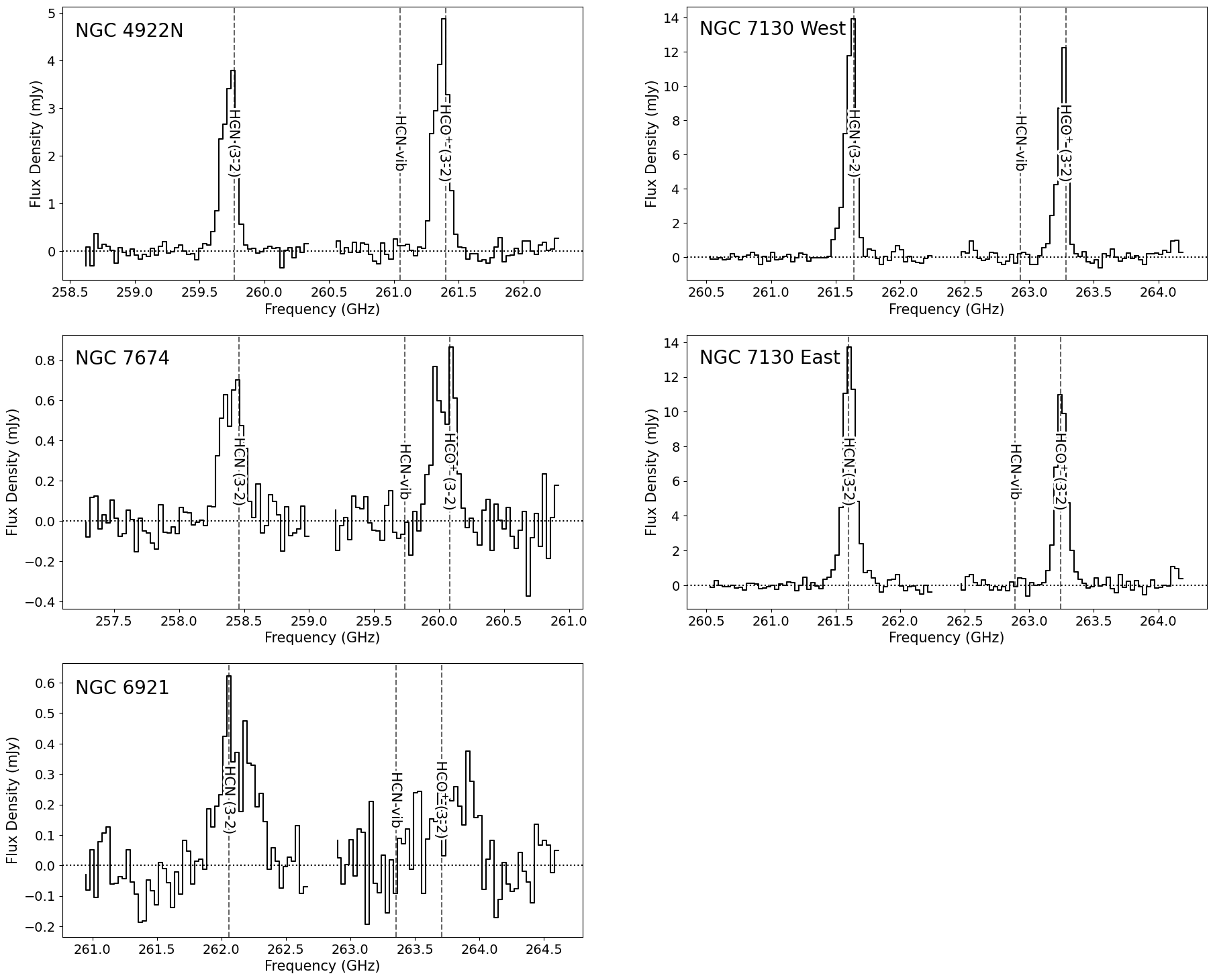}
\centering
\caption{Continuum-subtracted central spectra of the four CT AGN with new observations in our sample. The spectra were extracted from the peak of the 1-mm continuum using a beam-sized aperture (cyan ellipses in Figure 2). Redshifted frequencies of HCN (3-2), HCN-vib, and HCO$^+$ (3-2) are labeled by vertical dashed lines. Systemic velocities were set to match the peak of the HCN (3-2) line profile. Spectral line frequencies correspond to redshifts $z=0.02356$ (NGC 4922N), $z=0.01623$ (NGC 7130 West), $z=0.01639$ (NGC 7130 East), $z=0.01460$ (NGC 6921), and $z=0.02874$ (NGC 7674). }
\label{fig:spec}
\end{figure*}

\begin{deluxetable*}{ccccc}[!ht]
\tablecaption{Spectral line properties at the 1-mm continuum peak \label{tab:fluxes}}
\tablehead{ \colhead{Source} &  \colhead{Spectral Line} &  \colhead{Line Flux}  & \colhead{FWHM}  & \colhead{$\Sigma_{\text{line}}$}\\[-0.2cm] 
\colhead{} &  \colhead{} & \colhead{(Jy km s$^{-1}$)} & \colhead{(km s$^{-1}$)} & \colhead{($L_\odot$ pc$^{-2}$)}} 
\startdata
NGC 4922N & HCN (3-2) & 0.57 $\pm$ 0.04 & 148 $\pm$ 6 & 0.09 $\pm$ 0.01\\
& HCO$^{+}$ (3-2) & 0.69 $\pm$ 0.04 & 146 $\pm$ 6 & 0.11 $\pm$ 0.01 \\
& HCN-vib & $<$0.11 & -- & $<$0.02\\
\hline
NGC 7130 West & HCN (3-2) & 1.38 $\pm$ 0.05 & 102 $\pm$ 3  & 0.23  $\pm$ 0.02 \\
& HCO$^{+}$ (3-2) & 1.15 $\pm$ 0.05 & 95 $\pm$ 4 & 0.19  $\pm$ 0.02\\
& HCN-vib & $<$0.15 & -- & $<$0.02\\
NGC 7130 East & HCN (3-2) & 1.81 $\pm$ 0.06 & 110 $\pm$ 2  & 0.28 $\pm$ 0.03 \\
& HCO$^{+}$ (3-2) & 1.40 $\pm$ 0.06 & 119 $\pm$ 3 & 0.23 $\pm$ 0.02 \\
& HCN-vib & $<$0.17 & -- & $<$0.03\\
\hline
NGC 7674 & HCN (3-2) & 0.15 $\pm$ 0.04 & 226 $\pm$ 18& 0.03 $\pm$ 0.01 \\
& HCO$^{+}$ (3-2) & 0.16 $\pm$ 0.04 & 224 $\pm$ 25 & 0.04 $\pm$ 0.01\\
& HCN-vib & $<$0.13 & -- & $<$0.03\\
\hline
NGC 6921 & HCN (3-2) & 0.16 $\pm$ 0.07 & 357 $\pm$ 38 & 0.02 $\pm$ 0.01 \\
& HCO$^{+}$ (3-2) & 0.11 $\pm$ 0.07 & -- & 0.02 $\pm$ 0.01\\
& HCN-vib & $<$0.21 & -- & $<$0.03\\
\enddata
\tablecomments{Column (1): Source name. Column (2): Spectral line name. Column (3): Integrated flux at the 1-mm continuum peak in Jy km s$^{-1}$ measured with a beam-sized aperture. Uncertainties were estimated using the 1$\sigma$ RMS of spectrum and an assumed 10\% flux uncertainty for ALMA Band 6. Column (4): Full-width at half maximum of the spectral line profile. This was measured via a Gaussian fit to the line profile. Column (5): Surface brightness of spectral lines in L$_\odot$ / pc$^2$ measured over the beam area. Limits are provided at 3$\sigma$ confidence.}
\end{deluxetable*}
\vspace{-0.8cm}

We also note that NGC 7130 moment 0 maps show double-peaked emission consistent with the continuum emission. Intriguingly, the HCN (3-2) emission is brighter in the eastern peak whereas the western peak is brighter for both HCO$^{+}$ (3-2) and the 1-mm continuum. The source of the elevated HCN (3-2) emission around the western peak is briefly explored in Section \ref{sec:ratio}. 

We measure the fluxes of the HCN (3-2) and HCO$^+$ (3-2) spectral lines at the 1-mm continuum peak by extracting spectra with a beam-sized aperture (blue ellipses in Figure \ref{fig:mom0}). The continuum-subtracted central spectra are presented in Figure \ref{fig:spec}. We measure the Gaussian FWHM of the line profiles using the Astropy package \texttt{specutils}. The observed HCN (3-2) and HCO$^+$ (3-2) line profiles are relatively narrow and well-separated, so the fluxes were extracted over a velocity range with a size of twice the Gaussian FWHM. We note that a faint signal is detected at the expected red-shifted line center for HCO$^{+}$ (3-2) in NGC 6921. Assuming a FWHM equal to the HCN (3-2) line, we measure an integrated flux of 0.11 $\pm$ 0.07 Jy km s$^{-1}$ ($\sim$ 2$\sigma$) which we interpret as a marginally significant detection. 

We measure spectral line surface densities $\Sigma_{\text{line}}$ by dividing the line luminosity by the area of the beam-sized aperture (dimensions reported in Table \ref{tab:obs}). Here, we follow the prescription of \cite{Solomon2005} and define the line luminosity as
\begin{equation} \label{eq:lumin}
    \frac{L_{\text{line}}}{L_\odot} = \frac{1.04\times10^{-3}}{1+z}\left(\frac{S_{\text{line}}\Delta v}{\text{Jy km s}^{-1}}\right)\left(\frac{v_{\text{rest}}}{\text{GHz}}\right)\left(\frac{D_{\text{L}}}{\text{Mpc}}\right)^2,
\end{equation}
where $L_{\text{line}}$ is the line luminosity, $z$ is redshift, $S_{\text{line}}\Delta v$ is the velocity-integrated flux, $v_{\text{rest}}$ is the rest frequency of the spectral line, and $D_{\text{L}}$ is the luminosity distance. Spectral line properties and extracted flux measurements are reported in Table \ref{tab:fluxes}. For NGC 7130, measurements are reported for both the eastern and western continuum peaks. We find that the line properties are similar across the two peaks.

\vspace{1cm}
\subsection{Limits on HCN-vib emission}

Bright HCN-vib emission is undetected at the continuum peak of any of the four observed CT AGN (Figure \ref{fig:spec}). We also explore the regions surrounding the continuum peak, but HCN-vib emission is undetected. From the central beam spectra, we place 3$\sigma$ upper limits on the HCN-vib emission by assuming a boxcar line profile with a width equal to the FWHM of the HCN (3-2) line detection (presented in Table \ref{tab:fluxes}). We then apply Equation \ref{eq:lumin} and divide by the beam area to obtain a surface brightness upper limit. For example, the NGC 4922N spectrum has a mean 1$\sigma$ noise level of 0.25 mJy. Over a FWHM of 148 km s$^{-1}$, this corresponds to a $3\sigma$ flux upper limit of $<$ 0.11 Jy km s$^{-1}$, or a luminosity upper limit of $<$ 371 $L_\odot$ (Equation \ref{eq:lumin}). Measured over an aperture area of $\sim$ 70283 pc$^{2}$ (from the projected elliptical beam size in Table \ref{tab:obs}), we obtain a $3\sigma$ surface brightness upper limit of $\Sigma_{\text{HCN-vib}}<$ 0.02 $L_\odot$ pc$^{-2}$.

Following this procedure for all targets, we find that the $\Sigma_{\text{HCN-vib}}$ upper limits for the observed CT AGN fall below the 1 $L_\odot$ pc$^{-2}$ threshold for CONs at $\sim$100 pc-scales, indicating that they are not CONs. The measured HCN-vib upper limits are reported in Table \ref{tab:fluxes}. Note that we explore the effects of spatial resolution on these measurements in Section \ref{subsubsec:res}.

\subsection{Outflows and nuclear feedback}

Here, we search for molecular outflow signatures in the four observed CT AGN to assess the validity of a proposed evolutionary sequence that could link CONs and CT AGN. 

\subsubsection{Nuclear kinematics} \label{subsubsec:kinematics}

\begin{figure*}[!hp]
\centering
\includegraphics[width=0.7\textwidth]{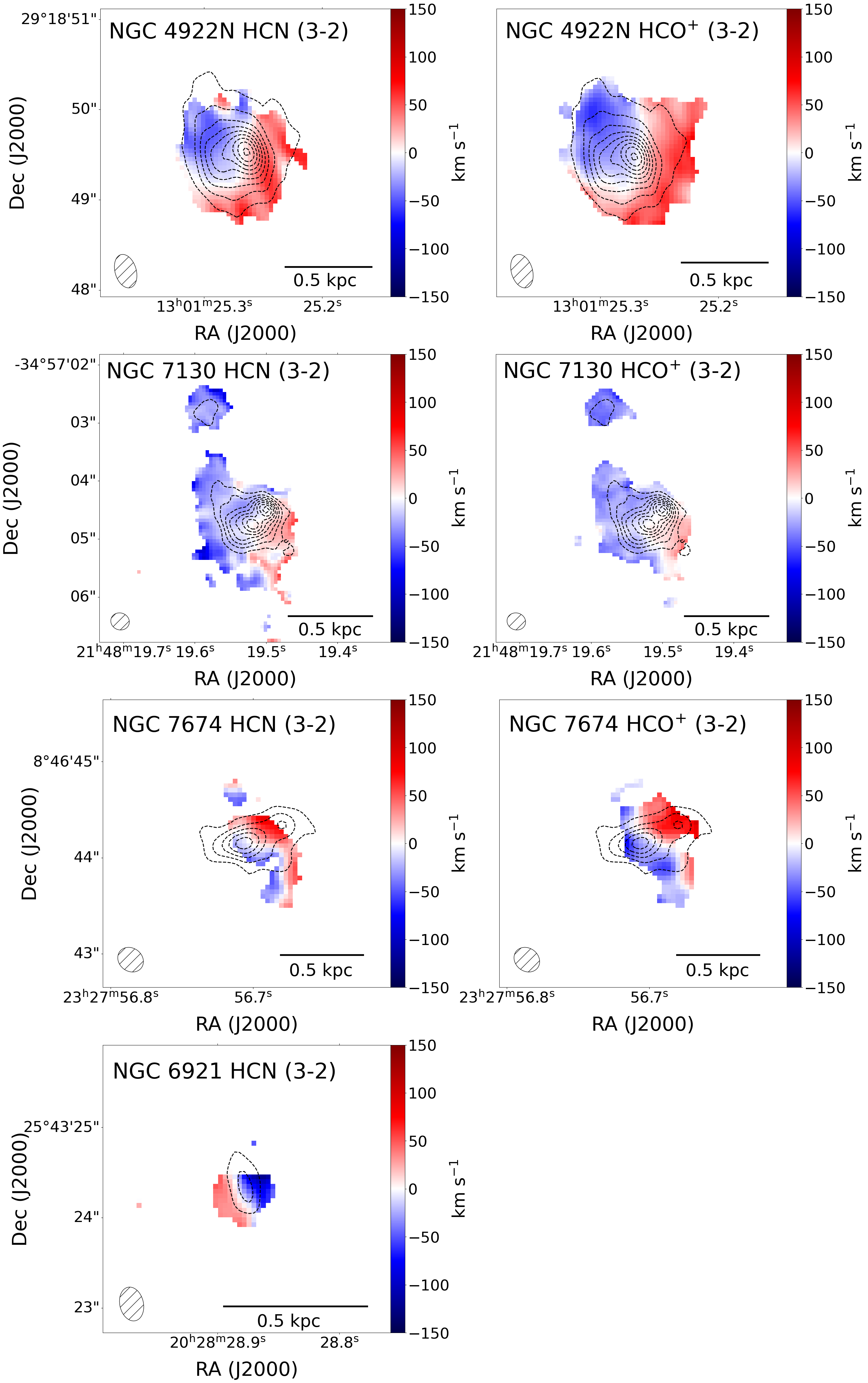}
\caption{Intensity-weighted velocity (Moment 1) maps of HCN (3-2) (left) and HCO$^+$ (3-2) (right). Each row corresponds with a source (from top to bottom:  NGC 4922N, NGC 7130, NGC 7674, NGC 6921). Moment 1 map extents were created by masking the data cube to where the total intensity (moment 0) $\geq 4\sigma$. 1-mm continuum contours (4$\sigma$ levels) are overlaid as black dashed lines. Note that NGC 6921 HCO$^+$ (3-2) emission is omitted because the emission features were faint and unresolved. \label{fig:mom1}}
\end{figure*}

\begin{figure*}[!hp]
\centering
\includegraphics[width=0.7\textwidth]{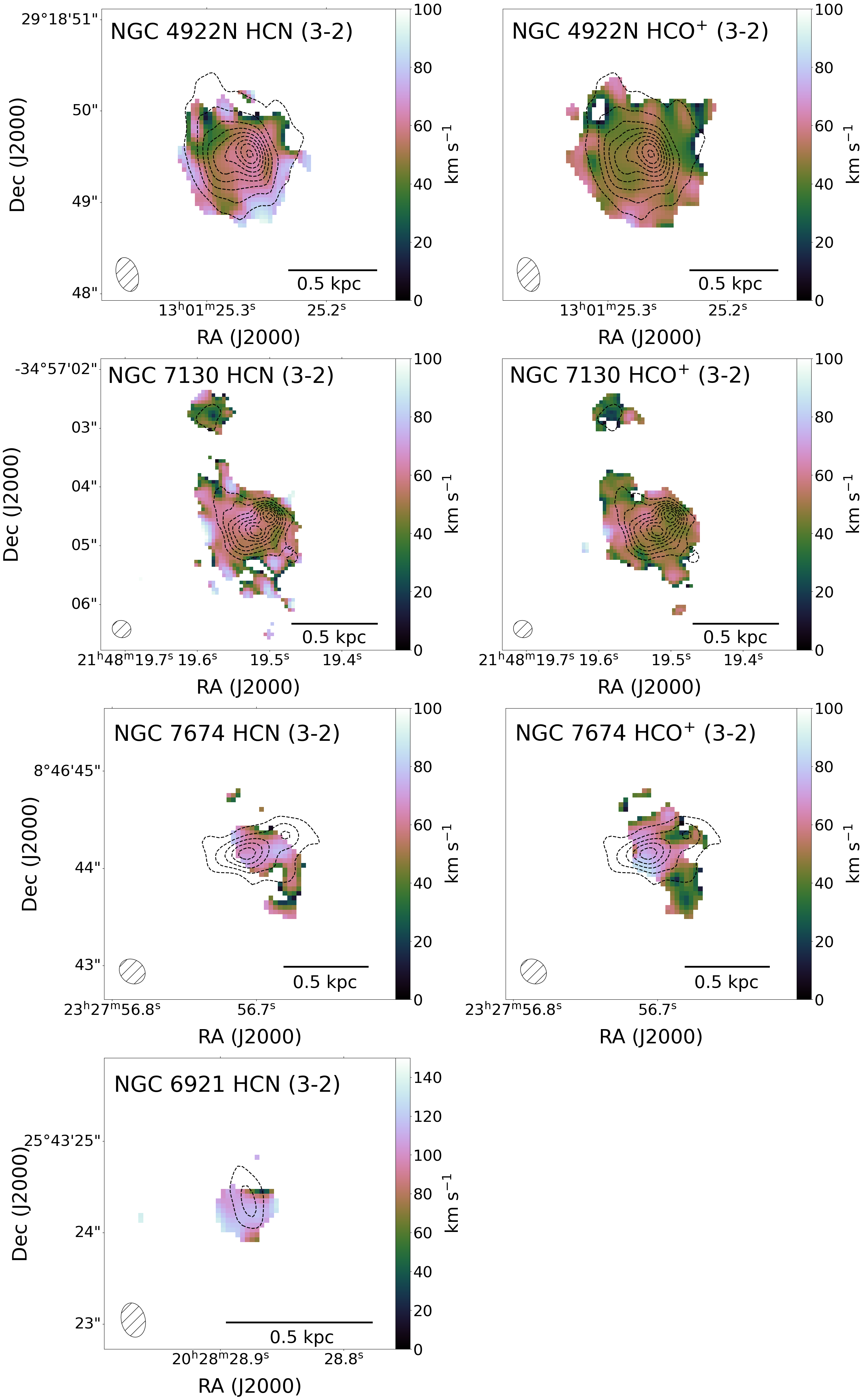}
\caption{Velocity dispersion (Moment 2) maps of HCN (3-2) (left) and HCO$^+$ (3-2) (right). Each row corresponds with a source (from top to bottom: NGC 4922N, NGC 7130, NGC 7674, NGC 6921). Moment 2 map extents were created by masking the data cube to where the total intensity (moment 0) $\geq 4\sigma$. 
1-mm continuum contours (4$\sigma$ levels) are overlaid as black dashed lines. Note that the NGC 6921 HCN (3-2) emission is shown with a wider dynamic range on the color-scale due to its higher velocity dispersions. NGC 6921 HCO$^+$ (3-2) emission is omitted because the emission features were faint and unresolved.  \label{fig:mom2}}
\end{figure*} 

We first analyze the nuclear kinematics of the four observed CT AGN and search for non-circular motions. Moment 1 (intensity-weighted velocity) and moment 2 (velocity dispersion) map are presented in Figure \ref{fig:mom1} and \ref{fig:mom2}, respectively. These images were made using the Python package \texttt{Spectral-Cube}, and the extents of the map were determined by masking the data cube to spatial regions where the total intensity (moment 0) is $\geq$ $3\sigma$. The moment 1 maps reveal velocity gradients approximately consistent with rotational motion (Figure \ref{fig:mom1}). Obvious peaks in velocity dispersion, however, are not present in any of the moment 2 maps for the four sources (Figure \ref{fig:mom2}). Note that for NGC 6921, HCO$^{+}$ (3-2) emission was omitted from this analysis because the emission was spatially unresolved and had a low signal-to-noise. The HCN (3-2) emission for NGC 6921, however, has the highest velocity dispersion among the four observed CT AGN.

\begin{figure*}[!ht]
    \centering
    \includegraphics[width=\textwidth]{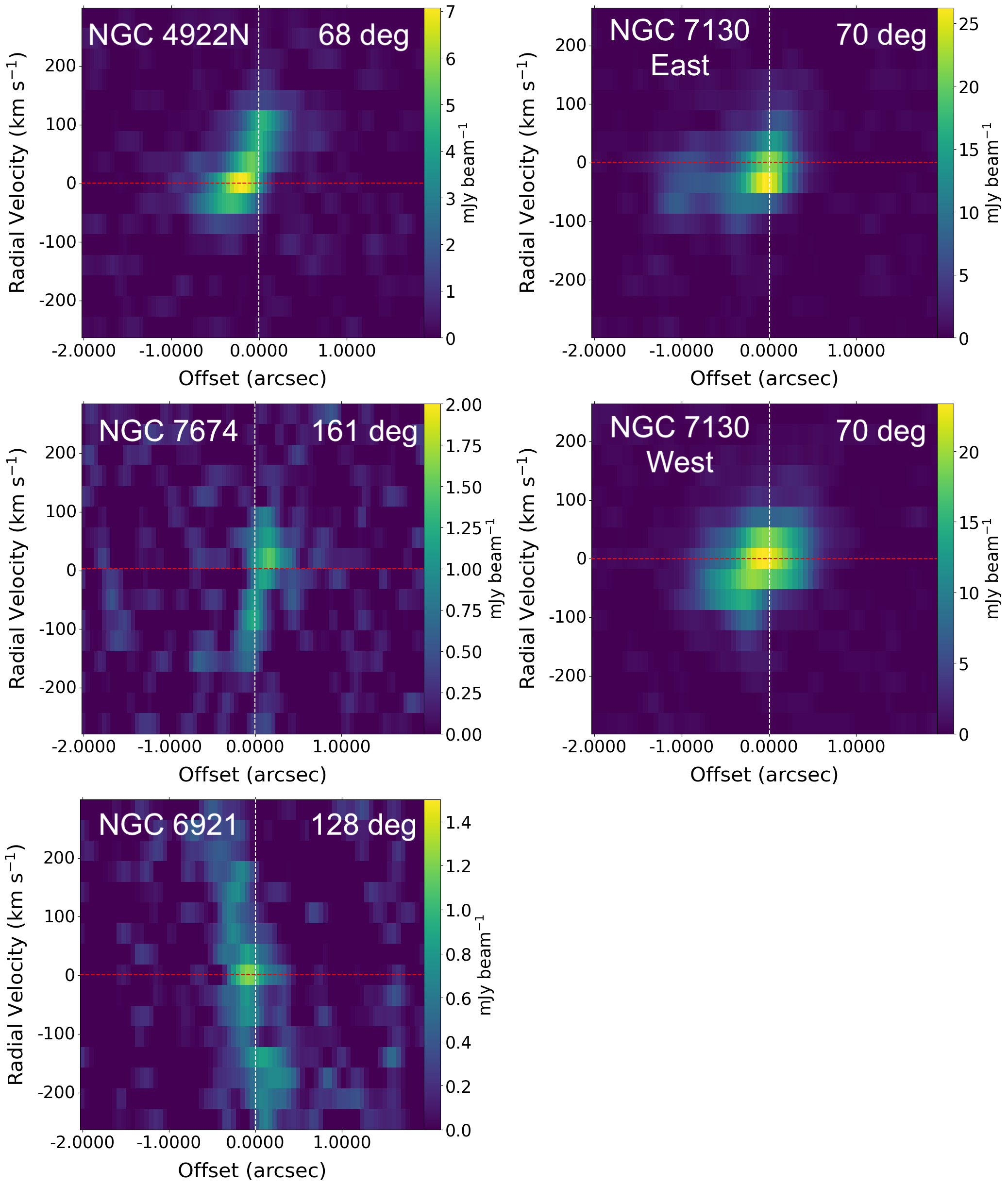}
    \caption{Position-velocity (PV) diagrams of the observed CT AGN. PV diagrams were created from a $\sim$2" long beam-width slice along the observed major axis of the velocity field. The position angles of the slices are shown in the top right of each plot. Red lines mark 0 km s$^{-1}$ correspond to the systemic velocity determined from the peak of the HCN (3-2) position. Vertical dashed white line marks the position of the 1-mm continuum peak. \label{fig:PV}}
    
\end{figure*}

Figure \ref{fig:PV} shows HCN (3-2) position-velocity (PV) diagrams taken along the observed major axis. PV slices were centered on the 1-mm continuum peak (see Figure \ref{fig:mom0}) with a $\sim$ 2$"$ slice length and an approximately beam-sized slice width. For the double-peaked source NGC 7130, we show separate PV diagrams with slices centered on each peak. Dashed red lines denote the systemic velocity determined from the peak of the HCN (3-2) line in the extracted spectra (see vertical dashed line in Figure \ref{fig:spec}). The white dashed line shows the location of the 1-mm continuum peak. For the western peak of NGC 7130, the emission feature extending to the left coincides with the position of the source's secondary eastern peak. Due to its low velocity nature, we conclude that this feature is not indicative of an outflow. In fact, obvious signs of non-circular motion are not seen in any  of the position-velocity diagrams. For all four sources, emission is contained within $\sim$ 1 arcsecond ($\sim$ $300-600$ pc) offsets from the continuum peak, and relative radio velocities do not exceed $\pm$ 200 km s$^{-1}$. With limited signs of outflowing molecular gas, the observed nuclear kinematics are instead consistent with the solid body components of the galaxies' rotation curves.  

\begin{figure*}[!ht]
    \centering
    \includegraphics[width=\textwidth]{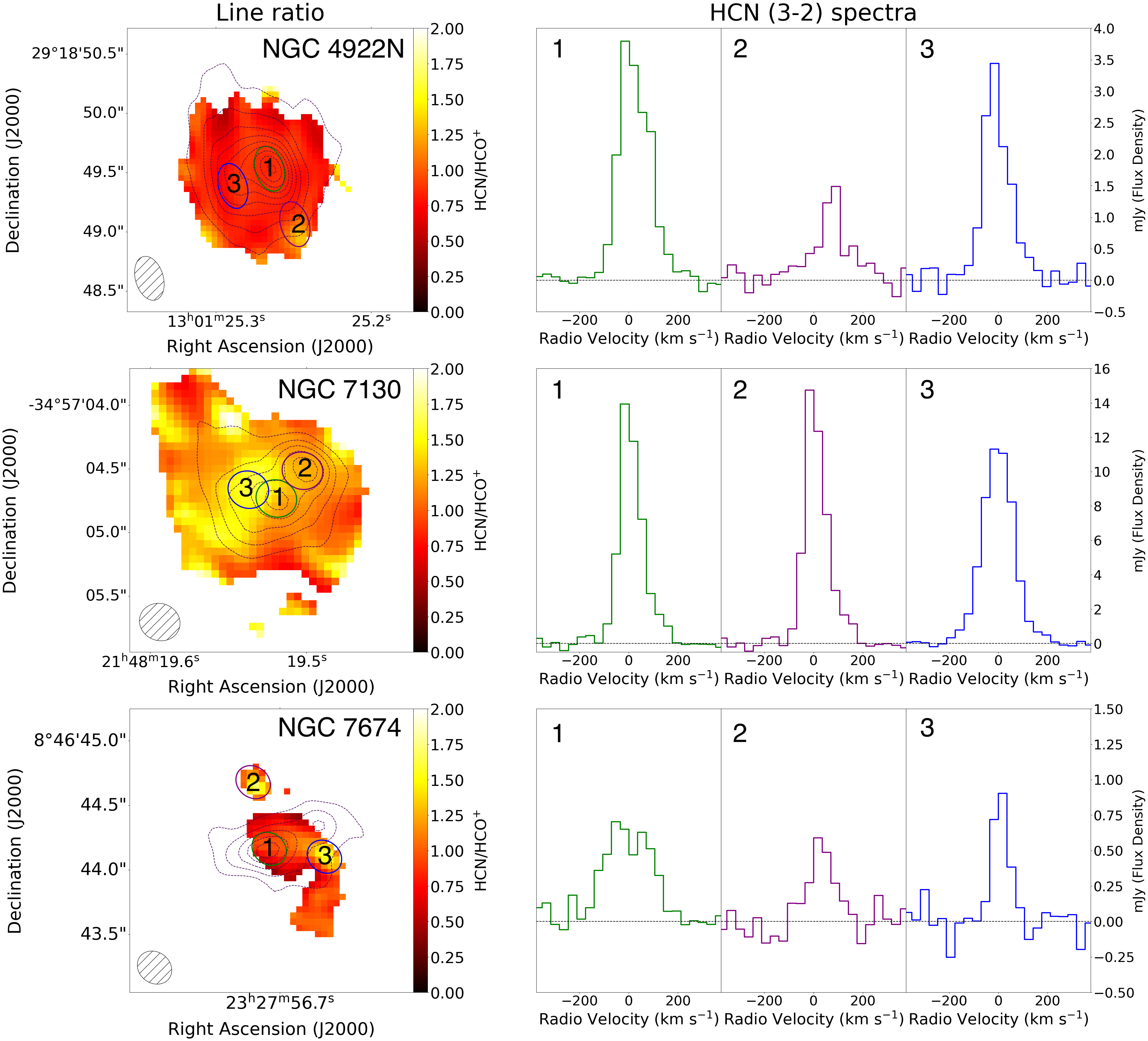}
    \caption{Left: Spatially resolved HCN (3–2)/HCO$^+$ (3-2) ratio maps of observed CT AGN. Right: Continuum-subtracted HCN (3-2) spectra extracted from beam-sized apertures. Beam-sized ellipses on the HCN (3–2)/HCO$^+$ (3-2) ratio map indicate spectral extraction regions. Dashed contours indicate 1-mm continuum emission. Spectra are numbered and shown in the same color as their corresponding ellipses. \label{fig:ratio}}
\end{figure*}

\subsubsection{\texorpdfstring{$HCN$/$HCO^+$ line ratio}{HCN/HCO+} }\label{sec:ratio}

In a further search for outflow signatures, we investigate spatial regions with elevated HCN/HCO$^+$ line ratios which can be driven by mechanical heating from outflows \citep{Izumi2015, Izumi2016}. Thus, spectral extractions from HCN-enhanced spatial regions could reveal outflow signatures via extended line wings. The left-hand panel of Figure \ref{fig:ratio} shows the spatially resolved $L_{\text{HCN (3-2)}}/$$L_{\text{HCO$^{+}$ (3-2)}}$ ratios of NGC 4922N, NGC 7130, and NGC 7674. NGC 6921 is again omitted from this analysis because the HCO$^+$ (3-2) emission was not spatially resolved and had a low signal-to-noise. We create this Figure by taking the ratio of the HCN (3-2) and HCO$^{+}$ (3-2) total intensity maps after masking out the regions where HCN (3–2) is not detected at $\geq3\sigma$. 

Beam-sized ellipses on the left-hand panel denote spatial regions from which HCN (3-2) spectra were extracted. These apertures are both numbered and colored to match their corresponding spectra which are displayed on the right-hand panel. For NGC 4922N, significant elevations in the line ratio are not observed. NGC 7130 and NGC 7674, however, reveal notable elevations in relative HCN (3-2) brightness. Though spectral extractions at these elevated spatial regions show asymmetries and line wings in the line profiles (notably, panel two for NGC 4922N and panel 3 for NGC 7130), the features are contained within $\sim\pm$ 200 km s$^{-1}$ of the systemic velocity, and obvious indicators of high velocity   molecular outflows are not present. We conclude that while low velocity ($\lesssim$ 200 km s$^{-1}$) non-circular motions may be present, fast outflows are not detected in regions with elevated HCN/HCO$^+$ ratios, and further investigation is necessary to determine the driver of the enhanced line ratio. 

\section{CT-AGN \& CONs  Connection in local U/LIRGs} \label{sec:CT_CON}

\subsection{Are CT AGN also CONs?}

\begin{figure}[!ht]
\includegraphics[width=0.5\textwidth]{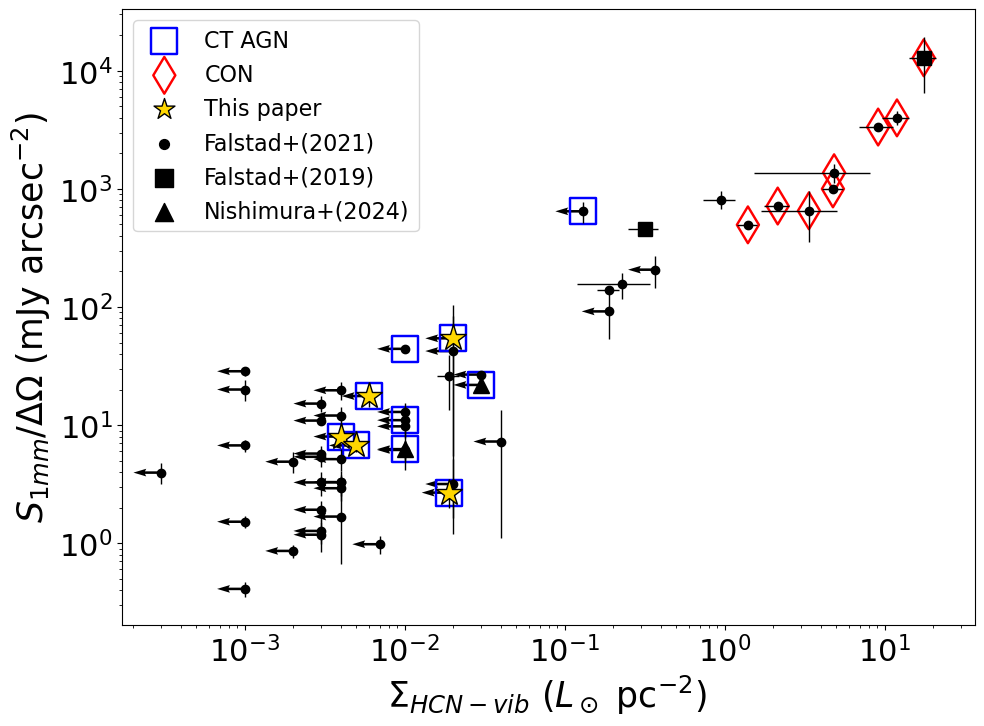}
\centering
\caption{1-mm continuum surface brightness as a function of HCN-vib surface brightness. Arrows indicate upper-limits. CONs and CT AGN are marked in red and blue respectively. Both axes are on a logarithmic scale. This Figure was adapted from Figure 10 in \cite{Falstad2021}. From their low HCN-vib surface brightnesses, we find that CT AGN are lacking CON-like infrared cores.}
\label{fig:CT_CON}
\end{figure}

Here, we combine our new continuum and HCN-vib measurements of four CT AGN with the literature measurements of five additional CT AGN. We can now compare the properties of the complete CT AGN sample with those of CONs. \cite{Falstad2021} found that CONs have both high HCN-vib surface brightnesses ($\Sigma_{\text{HCN-vib}} >$ 1 $L_\odot$ pc$^{-2}$) and 1-mm continuum surface brightnesses (S$_{\text{1-mm}}/\Omega \gtrsim$ 1000 mJy arcsec$^{-2}$), with both properties serving as tracers of high column density regions. In Figure \ref{fig:CT_CON}, we compare our complete sample of nine CT AGN to other local U/LIRGs (including CONs) by plotting $\Sigma_{\text{HCN-vib}}$ with respect to S$_{\text{1-mm}}/\Omega$ where measurements were available in the literature. All plotted sources were observed at $\sim 0.2-0.8"$ angular resolutions ($\lesssim100$ pc-scales; \citealt{Falstad2019,Falstad2021,Nishimura2024}; This work). Note that we plot measurements for both continuum peaks of NGC 7130 since the exact AGN position is unknown, resulting in 10 plotted values for our complete sample of nine CT AGN. We find that the CT AGN (shown in blue) have lower HCN-vib surface brightnesses and lower 1-mm continuum surface brightnesses than CONs (shown in red). 

In our comparison, we account for a methodological difference: \citet{Falstad2021} measured $\Sigma_{\text{HCN-vib}}$ across the continuum area defined by a Gaussian fit, whereas we report HCN-vib upper limits measured specifically with beam-sized aperture at the continuum peak in Table \ref{tab:fluxes}. While we elect to report upper-limits measured from the region where HCN-vib emission is most likely, we plot remeasured $\Sigma_{\text{HCN-vib}}$ upper limits in Figure \ref{fig:CT_CON}. These values were calculated using spectra extracted with a continuum-sized aperture (dimensions reported in Table \ref{tab:cont}). This ensures that the plotted $\Sigma_{\text{HCN-vib}}$ values and S$_{\text{1-mm}}/\Omega$ values were measured over the same area. The re-measured values are listed in Table \ref{tab:vib}. The plotted HCN-vib measurements of CT AGN in local U/LIRGs are listed in Table \ref{tab:vib}. 

\begin{deluxetable*}{cccc}[!ht]
\tablecaption{Continuum-area HCN-vib measurements of CT AGN in U/LIRGs  \label{tab:vib}}
\tablehead{\colhead{Source}  & \colhead{$\Sigma_{\text{HCN-vib}}$} & \colhead{$L_{\text{HCN-vib}}/L_{\text{IR}}$} & \colhead{Reference} \\[-0.2cm] 
\colhead{} & \colhead{($L_\odot$ pc$^{-2}$)} & \colhead{($10^{-8}$)} & \colhead{}}
\startdata
NGC 1068 & $<0.13$ & $<$0.004 & \cite{Falstad2021} \\
NGC 4922N & $<0.004$ & $<0.153$ & This work \\
13120-5453 & $<0.01$ & $<0.108$ & \cite{Falstad2021} \\
NGC 5135 & $<0.01$ & $<0.085$ & \cite{Falstad2021} \\
NGC 6240N & $<0.01$ & $<0.128$ & \cite{Nishimura2024} \\
NGC 6240S & $<0.03$ & $<0.238$ & \cite{Nishimura2024} \\
NGC 6921 & $<0.03$ &  $<0.249$ & This work \\
NGC 7130 E & $<0.02$ & $<$0.082 & This work \\
NGC 7130 W & $<0.006$ & $<$0.085 & This work \\
NGC 7674 & $<0.005$ &  $<0.120$ & This work 
\enddata
\tablecomments{Column(1): Source name.  Column (2): HCN-vib surface brightnesses and upper-limits in $L_\odot$ / pc$^2$. Limits are provided at 3$\sigma$ confidence. Column (3):  HCN-vib luminosity to IR luminosity ratio. Limits are provided at 3$\sigma$ confidence. Column (4): Literature from which HCN-vib measurements were obtained. }
\end{deluxetable*}

Figure \ref{fig:CT_CON} reveals that the HCN-vib surface brightnesses and 1-mm continuum surface brightnesses of CT AGN do not overlap with that of the CONs. In fact, all \textit{NuSTAR} hard X-ray-confirmed CT AGN have $\Sigma_{\text{HCN-vib}} <$ 1 $L_\odot$ pc$^{-2}$ \citep{Martin2016, Falstad2019, Falstad2021, Nishimura2024}, translating to a CONs detection rate of 0$^{+17}_{-0}$\% in CT AGN. Here, the 1-$\sigma$ confidence intervals were estimated using the beta distribution quantile technique which assigns a probability density function to an assumed binomial population. This method is outlined in \cite{Cameron2011}. 

The observed CT AGN also lack significant self-absorption features in the HCO$^+$ (3-2) and HCN (3-2) line profiles (Figures \ref{fig:spec} and \ref{fig:ratio}). CONs, on the other hand, have double-peaked spectral lines (e.g. IC 860, \citealt{Aalto2019}) that indicate the absorption of photons by an enshrouding layer of cooler gas \citep{Aalto2015}. With the exception of NGC 7674, the observed spectral lines are not double-peaked (see Figures \ref{fig:spec} and \ref{fig:ratio}), suggesting that line-of-sight column densities of these CT AGN may be smaller than those of typical CONs.

CT AGN and CONs may also differ in their infrared spectral energy distributions (SEDs), but we cannot fully explore this property due to a sample selection bias. \citet{Falstad2021} found that CONs have low IRAS flux ratios at 25 $\mu$m and 60 $\mu$m ($f_{25}/f_{60}$), but the surveyed sample was biased towards sources with cooler infrared SEDs. Our sample of hard X-ray-confirmed CT AGN include both warm ($f_{25}/f_{60} >0.2$) and cool ($f_{25}/f_{60}<0.2$) sources (see Table \ref{tab:sample}), revealing a different $f_{25}/f_{60}$ distribution than the CONs detected in CON-quest. While this could indicate a significant difference in the properties of CT AGN and CONs, we cannot precisely compare their $f_{25}/f_{60}$ distributions until an unbiased search for warm CONs is conducted. 

\subsubsection{Impact of Spatial Resolution} \label{subsubsec:res}

The $\Sigma_{\text{HCN-vib}}$ upper limits in Table \ref{tab:vib} demonstrate that CT AGN in U/LIRGs generally do not have HCN-vib detections at $\sim$100 pc-scales. While this could indicate that the CT AGN have a weaker mid-infrared radiation field than CONs (and thus cannot excite HCN molecules into vibrationally excited states), the inferred $\Sigma_{\text{HCN-vib}}$ upper limits could be affected by insufficient spatial resolution. We consider here that CT AGN in U/LIRGs may have smaller HCN-vib emission regions than the spatial scales probed in this work. This is motivated by previous studies of the CT AGN NGC 1068 that detected high surface brightness HCN-vib emission at $\sim$ 2 pc spatial resolution ($\Sigma_{\text{HCN-vib}} = 1.6$ $L_\odot$ pc$^{-2}$; \citealt{Imanishi2016})  but a non-detection at $\sim$ 15 pc-scales ($\Sigma_{\text{HCN-vib}} < 0.13$ $L_\odot$ pc$^{-2}$; \citealt{Falstad2021}).

We evaluate the effects of spatial resolution on $\Sigma_{\text{HCN-vib}}$ by approximating the HCN-vib emission region as a single-point source for each CT AGN. We can then estimate the aperture size necessary to obtain an HCN-vib surface brightness measurement that is consistent with the CONs criterion ($\Sigma_{\text{HCN-vib}} \geq 1$ $L_\odot$ pc$^{-2}$). Note that this approximation assumes that the line luminosity remains constant, and only the radius of the aperture is adjusted (i.e. the emission region size).  If we assume a HCN-vib line luminosity equal to the 3$\sigma$ upper limit (see Table \ref{tab:vib}), then the radius upper limit ranges from $r<$ $4-26$ pc and a median of $r<$ 16 pc to have a sufficiently high $\Sigma_{\text{HCN-vib}}$. A 1$\sigma$ line luminosity requires an even more compact core, with $r<$ $2-15$ pc and a median upper limit of $r<$ 10 pc. We emphasize that our empirical fiducial value of $\Sigma_{\text{HCN-vib}} > 1$ $L_\odot$ pc$^{-2}$ for CONs was selected for measurements taken over a region with $r>5$pc \citep{Falstad2021}, and smaller radii are inconsistent with the observed properties of CONs.

We also consider the fiducial value $L_{\text{HCN-vib}}/L_{\text{IR}} > 10^{-8}$ which was used by early studies of CONs (e.g., \citealt{Aalto2015, Falstad2019}). Though this luminosity ratio is a less robust metric because the IR emission can have contributions from non-nuclear regions of the galaxy, it nevertheless offers a criterion that is less influenced by the systematic effects of spatial resolution. In Table \ref{tab:vib}, we report $L_{\text{HCN-vib}}/L_{\text{IR}}$ ratio measurements of the CT AGN studied in this work. We find that the CT AGN are not CONs with respect to both the $\Sigma_{\text{HCN-vib}}$ criterion and the $L_{\text{HCN-vib}}/L_{\text{IR}}$ criterion. 

Though we cannot eliminate the possibility that our sample of CT AGN has parsec-scale CON-like HCN-vib emission regions, the non-detections of CONs across both fiducial values combined with the clear divergence in the HCN-vib properties at $\sim100$ pc scales demonstrate that CONs and CT AGN are distinct phenomena in U/LIRGs. We therefore conclude that the limited spatial resolution does not significantly affect the broader conclusion of this work. 

\subsection{Evolutionary link: do CT AGN evolve from CONs?}

Although the properties of CT AGN deviate from those of CONs at $\sim100$ pc-scales, their gas kinematic features could provide insight into their potential evolutionary link to CONs. CONs host collimated molecular outflows with maximum velocities ranging from 300 km s$^{-1}$ (ESO 320-G030; \citealt{PS2016}) to 840 km s$^{-1}$ (Arp 220W; \citealt{BM2018}). Here, we explore the proposed feedback-driven evolutionary sequence in which CT AGN would have faster and more evolved molecular outflows than the collimated features found in CONs \citep{GA2017b, Falstad2019, Falstad2021}.

Based on the line luminosities of known HCN outflows in CONs, we confirm that our sensitivities are sufficient to detect similar outflows in our observed CT sample, if they are present. Specifically, the HCN (3-2) outflow in ESO 320-G030 \citep{Gorski2024} would have been detected at a $8-28\sigma$ level. Likewise, the HCN (1-0) outflow in Arp 220W \citep{BM2018} would have been detected at a $3-11\sigma$ level. The latter calculation assumes an HCN (3-2) to HCN (1-0) luminosity ratio of 0.5 \citep{Aalto2015b, Gorski2024}.

Despite this potential for outflow detection, we find no signs of high velocity ($>$ 200 km s$^{-1}$) non-circular motion in the HCN (3-2) emission of the four observed CT AGN. Such non-detections are inconsistent with the proposed evolutionary sequence. Examining the literature on our combined sample of nine CT AGN, however, we find that a significant fraction of them do host fast molecular outflows. Specifically, high velocity ($>$ 500 km s$^{-1}$) CO outflows have been detected in NGC 6240 (a merger system containing two CT AGN, NGC 6240N and NGC 6240S; \citealt{Feruglio2013, Treister2020}). Dense molecular outflows have also been observed in NGC 1068 and IRAS 13120-5453 with velocities of $\sim 300$ km s$^{-1}$ \citep{GB2014, Privon2017, Lutz2020}. Of the nine hard X-ray-confirmed CT AGN, we find that at least 44\% (4/9) host molecular outflows.\footnote{We acknowledge that it is unclear whether all of these molecular outflows are powered by an AGN. Specifically, NGC 6240 hosts a multi-component outflow with both a likely starburst-driven component and a likely AGN-driven component \citep{MS2018}. This ambiguity makes it difficult to assess whether the outflow properties of NGC 6240 agree with the proposed evolutionary link between CONs and CT AGN.} 

According to the proposed sequence, these molecular outflows may have evolved from the collimated outflows of CONs. We estimate the ages of the outflows to test this theory. We find that the collimated outflows have short predicted dynamical times ($\sim$ 0.2 Myr; \citealt{PS2016,BM2018, Falstad2018, Lutz2020}), whereas the molecular outflows in CT AGN have comparatively older outflow ages ($\gtrsim 1$ Myr; \citealt{GB2014, Lutz2020, Feruglio2013, Treister2020}). The presence of these older, well-developed molecular outflows coupled with weak HCN-vib emission makes these CT AGN qualitatively consistent with the proposed sequence of evolution. 

Detections of large-scale ionized outflows in CT AGN may also be consistent with the proposed sequence, considering that the gas in a previous CON-like molecular outflow may have been ionized by the central source. The larger physical scales of ionized outflows often indicate that they are older than the assumed duration of the active feedback phase ($\sim$ 1 Myr), possibly associating them with a previous epoch of CON activity. For example, ionized outflows with estimated expansion times of 3$-$11 Myr have been observed in some CT AGN in U/LIRGs (e.g., NGC 7130 and NGC 6240; \citealt{Comeron2021, MS2018}).

Given that some CT AGN host well-developed outflows while others do not, it is possible that there are multiple evolutionary paths for CT AGN, with some sources evolving from CONs as previously theorized and others having never entered a CON phase. We find evidence of this theory in the fact that the estimated lifetime of the CON phase is significantly shorter ($\sim$ 1 Myr; \citealt{Aalto2019}) compared to that of the CT AGN phase ($\sim100$ Myr). The ratio of these estimated timescales implies that even if all CONs evolve into CT AGN, they would only account for 1\% of the CT AGN population. This further suggests that only a small subset of CT AGN is likely related to CONs. 

In an idealized scenario, the presence of outflows could be used to distinguish between CT AGN that evolved from CONs and those that did not. Unfortunately, CT AGN without active molecular outflows could still have evolved from CONs if the outflow activity is episodic. That is, the outflow may have become inactive after entering the CT AGN phase, resulting in a non-detection. This model is supported by the ample theoretical and observational evidence of episodic outflow activity in local U/LIRGs which then make it reasonable to infer similar mechanisms in CONs and CT AGN. For example, fast blow-out phases are predicted by the radiative feedback model ($\sim$ 1 Myr, \citealt{Hopkins2008, Ricci2017nat, Ricci2022, Yutani2022}), implying short-lived active feedback mechanisms. Outflow `fossils' from previous epochs of outflow activity have also been detected in local U/LIRGs (e.g, \citealt{Janssen2016, Lutz2020, HC2020}). A complete search for these outflow relics in CONs and CT AGN, however, has not been conducted and will be necessary to fully evaluate the model.

The possibility of episodic outflow activity as well as multiple evolutionary paths complicates current and future studies of CON evolution. The proposed sequence, however, necessitates the existence of nuclei that are in the process of transitioning out of a CON phase and into a CT AGN phase. Direct detection of these intermediate-stage sources as well as the characterization of their outflow properties and their HCN-vib emission will inform and constrain the evolutionary model. The future directions and observational limitations of this research are discussed in Section \ref{subsubsec:extremeCT} below.

\subsection{Sample selection bias - Heavily-CT AGN} \label{subsubsec:extremeCT}

Here, we consider the effects of a potential sample selection bias on the low measured rate of occurrence of CONs in U/LIRGs that host CT AGN. 

\begin{figure}[!ht]
\includegraphics[width=0.5\textwidth]{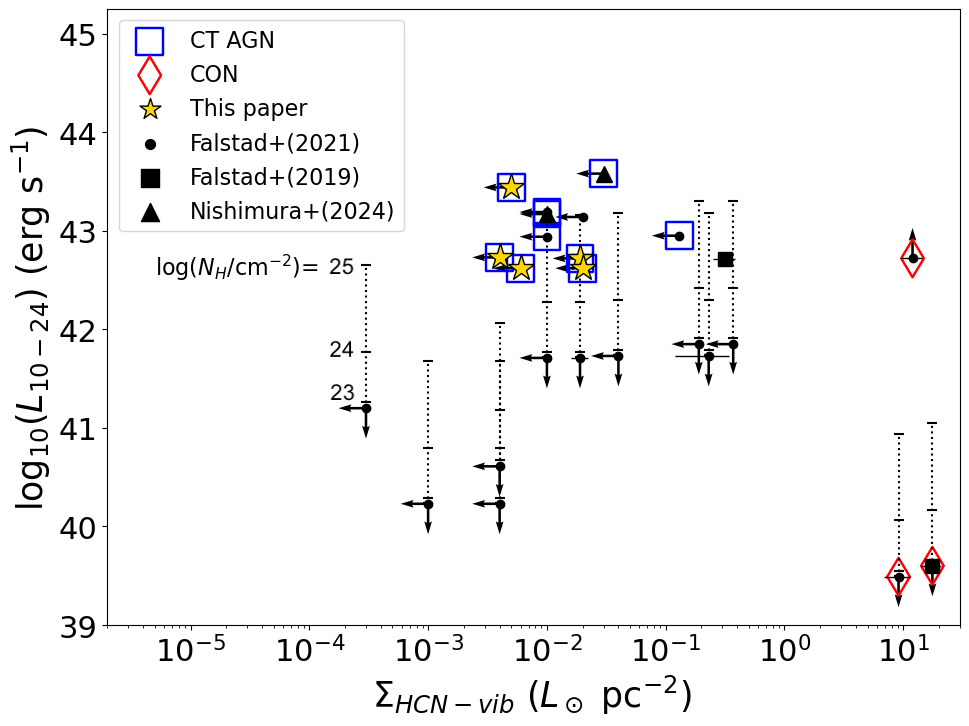}
\centering
\caption{Extinction-corrected hard X-ray luminosity vs. HCN-vib surface brightness of CT AGN and CONs. CT AGN are marked in blue, and CONs are in red. Arrows indicate upper-limits. Dotted lines show $N_{\text{H}}$ corrections for X-ray luminosity upper-limits, with the upper limits shifting to the subsequent tick for every order of magnitude increase in line-of-sight column density. The limited number of hard X-ray detections of CONs (shown in red) suggests that if CONs host CT AGN, they are likely undetectable at hard X-ray wavelengths. }
\label{fig:X_vib}
\end{figure}

Examining the intrinsic properties of the CT AGN in our sample, we find limited spread in their extinction-corrected hard X-ray luminosities. This is consistent with findings of \citet{Ricci2021} that reported that obscured AGN tend to have higher intrinsic X-ray luminosities than unobscured AGN. Figure \ref{fig:X_vib} is a plot of the 10$-$24 keV intrinsic X-ray luminosities from the literature \citep{Teng2015,Puccetti2016, Gandhi2017, Ricci2017a,Ricci2017c, Oda2018,Yamada2020,Ricci2021} with respect to the HCN-vib surface brightnesses of local U/LIRGs (this work; \citealt{Falstad2019, Falstad2021, Nishimura2024}). Arrows indicate upper limits for HCN-vib surface brightness and X-ray luminosity. Dotted lines are $N_{\text{H}}$ corrections for undetected but possibly obscured sources. Note that the low number of data points from \citet{Falstad2021} is due to the limited number of published hard X-ray measurements for the sample. 

Though hard X-ray studies offer a reliable and direct means of identifying AGN, X-ray studies of CONs have been unable to detect AGN (as demonstrated by CONs non-detections/upper-limits in Figure \ref{fig:X_vib}). Despite these AGN non-detections in the X-rays, a starburst scenario is also uncertain as it would require O-star dominated IMFs at 10$-$100 pc-scales (\citealt{Aalto2019}). An alternative interpretation, one that is consistent with our proposed evolutionary model, is that CONs host reflection-dominated AGN with large gas columns and covering factors that significantly reduce X-ray detections. In fact, \cite{Ricci2017b} finds that the observed rate of `heavily-CT' AGN ($N_{\text{H}}>10^{25}$ cm$^{-2}$) in the local Universe is significantly lower than the intrinsic rate determined from the inferred $N_{\text{H}}$ distribution. Models further suggest that a high fraction of AGN are X-ray obscured \citep{Worsley2005, Carroll2023, Giovanni2024}. Indeed, with the exception of NGC 4922N, there is an absence of sources with line-of-sight column densities $N_{\text{H}}>10^{25}$ cm$^{-2}$ in our sample. Thus, our study does not eliminate the possibility that CONs are an extreme subset of CT AGN ($N_{\text{H}}>10^{25}$ cm$^{-2}$) that are non-detectable at X-ray wavelengths due to attenuation. 

If CONs are driven by heavily-CT AGN, a source of interest would be Arp 220W, a known CON \citep{Aalto2015, Martin2016, Sakamoto2021} with an estimated column density as high as $N_{\text{H}} \approx 10^{26}$ cm$^{-2}$ \citep{Scoville2017}. \textit{NuSTAR} observations detect X-ray emission that is consistent with only a starburst, but due to its high column density, a deeply buried AGN may be present in Arp 220W \citep{Teng2015}. In fact, recent studies have uncovered evidence of possibly AGN-driven outflows \citep{Varenius2016, BM2018}. The AGN nature of the CON Arp 220W has been fiercely debated due to this mixed evidence, and there is increased urgency to identify new techniques to detect AGN in heavily obscured environments. 

A promising solution is the recently identified mm-X-ray relation \citep{Kawamuro2022, Ricci2023} that has the potential to confirm (or dispute) the presence of an AGN in Arp 220W and other heavily obscured nuclei. The tight correlation between AGN X-ray luminosity and mm-wave luminosity \citep{Kawamuro2022, Ricci2023} suggests that flat spectrum ($\alpha_{\text{mm}} \approx 0 $) mm-wave emission could be used to derive the intrinsic X-ray luminosities of heavily obscured AGN (and subsequently their bolometric luminosities). Since mm-waves are almost unaffected by dust-extinction until $N_{\text{H}} \sim 10^{26}$ cm$^{-2}$, mm-wave observations may prove to be essential in the search for these `missing' heavily-CT AGN. 

We also note that there is a need to search for the CONs in CT AGN outside of the U/LIRGs in the GOALS sample (for example, CT AGN in post-starburst galaxies) which would provide a statistic on the prevalence of CONs in a more complete sample of CT AGN.  The apparent strong infrared luminosity-dependence of the CON phenomena \citep{Falstad2021}, however, makes it unlikely that such a search would be fruitful.

\section{Conclusion} \label{sec:concl}

We study the prevalence of CONs in CT AGN in local U/LIRGs by analyzing ALMA Band 6 HCN-vib observations of four local U/LIRGs hosting a CT AGN, combined with five additional sources with HCN-vib literature data. We study this combined sample of nine CT AGN with HCN-vib measurements to analyze the potential connection between CT AGN and CONs in local U/LIRGs. Our results are as follows.

\begin{enumerate}[topsep=8pt,itemsep=6pt,partopsep=6pt, parsep=6pt]
    \item Using the fiducial value of $\Sigma_{\text{HCN-vib}} \geq$ 1 $L_\odot$ pc$^{-2}$, we find a CON detection rate of 0$^{+17}_{-0}$\% in hard X-ray-detected CT AGN in local U/LIRGs (see Figure \ref{fig:CT_CON}). These CT AGN also have significantly fainter 1-mm continuum surface brightnesses compared to CONs. This reveals that X-ray-confirmed CT AGN do not host CONs. 

    \item While all known CONs have shown evidence of molecular outflows \citep{Falstad2021}, we do not find any evidence of significant high velocity non-circular motion in the dense gas emission of four observed CT AGN (Figures \ref{fig:PV} and \ref{fig:ratio}). We note, however, that fast molecular outflows have been reported in the literature for 4/9 (44\%) of the CT AGN in our sample. The fact that some CT AGN in U/LIRGs host molecular outflows while others have non-detections suggest that there may be multiple evolutionary paths for CT AGN (with some evolving from a prior CONs phase, and others not). Alternatively, this could be indicative of a relatively short outflow duty cycle compared to the timescale of the CT phase of AGN evolution.

\end{enumerate}

%



\begin{acknowledgments}
\section*{Acknowledgements}
This research was supported by NASA awards 80NSSC21K1177 and
80NSSC22K0064. This paper makes use of the following ALMA data: ADS/JAO.ALMA\#2017.1.00598.S. ALMA is a partnership of ESO (representing its member states), NSF (USA) and NINS (Japan), together with NRC (Canada), MOST and ASIAA (Taiwan), and KASI (Republic of Korea), in cooperation with the Republic of Chile. The Joint ALMA Observatory is operated by ESO, AUI/NRAO and NAOJ. The National Radio Astronomy Observatory is a facility of the National Science Foundation operated under cooperative agreement by Associated Universities, Inc. M.A.J. gratefully acknowledges financial support for this research by the Fulbright U.S. Student Program, which is sponsored by the U.S. Department of State and US-Chile Fulbright Commission. Its contents are solely the responsibility of the author and do not necessarily represent the official views of the Fulbright Program, the Government of the United States, or the US-Chile Fulbright Commission. ET and FEB gratefully acknowledge funding from ANID programs FONDECYT Regular 1200495, CATA-BASAL FB210003, and Millennium Science Initiative Programs NCN19\_058. CR acknowledges support from Fondecyt Regular grant 1230345 and ANID BASAL project FB210003. JSG thanks the University of Wisconsin  College of Letters and Science and Macalester College for partial support of his research on CONs. 
\end{acknowledgments}

\software{ipython \citep{ipython}, numpy \citep{numpy}, Astropy \citep{Astropy}, and Spectral-Cube \citep{SpectralCube}.}

\bibliography{biblio}{}
\bibliographystyle{aasjournal}

\end{document}